\documentclass[aps,pra,twocolumn,showpacs,amsmath,superscriptaddress,floatfix,nofootinbib]{revtex4-1}
\usepackage[subnum]{cases}
\usepackage[utf8]{inputenc}
\usepackage{graphicx}
\usepackage{dcolumn}
\usepackage{hyperref}

\newcommand*{\figurewidth}{\columnwidth}
\newcommand*{\etal}{\emph{et al.\ }}
\newcommand*{\eref}[1]{Eq.~(\ref{#1})}
\newcommand*{\erefs}[3]{Eqs.~(\ref{#1})#3(\ref{#2})}
\newcommand*{\tref}[1]{Tab.~\ref{#1}}
\newcommand*{\sref}[1]{Sec.~\ref{#1}}
\newcommand*{\fref}[1]{Fig.~\ref{#1}}

\begin{document}

\title{Absolute rate coefficients for photorecombination and electron-impact ionization of magnesium-like iron ions from measurements at a heavy-ion storage ring}
\author{D.~Bernhardt} 
\affiliation{Institut f\"{u}r Atom- und Molek\"{u}lphysik, Justus-Liebig-Universit\"{a}t Giessen, Leihgesterner Weg 217, D-35392 Giessen, Germany}
\author{A.~Becker}
\affiliation{Max-Planck-Institut f\"{u}r Kernphysik, Saupfercheckweg 1, D-69117 Heidelberg, Germany}
\author{M.~Grieser}
\affiliation{Max-Planck-Institut f\"{u}r Kernphysik, Saupfercheckweg 1, D-69117 Heidelberg, Germany}
\author{M.~Hahn}
\affiliation{Columbia Astrophysics Laboratory, Columbia University, New York, NY 10027, USA}
\author{C.~Krantz}
\affiliation{Max-Planck-Institut f\"{u}r Kernphysik, Saupfercheckweg 1, D-69117 Heidelberg, Germany}
\author{M.~Lestinsky}
\affiliation{GSI Helmholtzzentrum f\"{u}r Schwerionenforschung, Planckstra{\ss}e 1,
D-64291 Darmstadt, Germany}
\author{O.~Novotn\'y}
\affiliation{Columbia Astrophysics Laboratory, Columbia University, New York, NY 10027, USA}
\author{R.~Repnow}
\affiliation{Max-Planck-Institut f\"{u}r Kernphysik, Saupfercheckweg 1, D-69117 Heidelberg, Germany}
\author{D.~W.~Savin}
\affiliation{Columbia Astrophysics Laboratory, Columbia University, New York, NY 10027, USA}
\author{K.~Spruck}
\affiliation{Institut f\"{u}r Atom- und Molek\"{u}lphysik, Justus-Liebig-Universit\"{a}t Giessen, Leihgesterner Weg 217, D-35392 Giessen, Germany}
\affiliation{Max-Planck-Institut f\"{u}r Kernphysik, Saupfercheckweg 1, D-69117 Heidelberg, Germany}
\author{A.~Wolf}
\affiliation{Max-Planck-Institut f\"{u}r Kernphysik, Saupfercheckweg 1, D-69117 Heidelberg, Germany}
\author{A.~M\"{u}ller}
\affiliation{Institut f\"{u}r Atom- und Molek\"{u}lphysik, Justus-Liebig-Universit\"{a}t Giessen, Leihgesterner Weg 217, D-35392 Giessen, Germany}
\author{S.~Schippers} \email{Stefan.Schippers@physik.uni-giessen.de}
\affiliation{Institut f\"{u}r Atom- und Molek\"{u}lphysik, Justus-Liebig-Universit\"{a}t Giessen, Leihgesterner Weg 217, D-35392 Giessen, Germany}

\begin{abstract}
Rate coefficients for photorecombination (PR) and cross sections for electron-impact ionization (EII) of Fe$^{14+}$ forming Fe$^{13+}$ and Fe$^{15+}$, respectively, have been measured by employing the electron-ion merged-beams technique at a heavy-ion storage ring. Rate coefficients for PR and EII of Fe$^{14+}$ ions in a plasma are derived from the experimental measurements. Simple parametrizations of the experimentally derived plasma rate coefficients are provided for use in the modeling of photoionized and collisionally ionized plasmas. In the temperature ranges where Fe$^{14+}$ is expected to form in such plasmas  the latest theoretical rate coefficients of Altun \etal [Astron.\ Astrophys.\ \textbf{474}, 1051 (2007)] for PR and of Dere [Astron.\ Astrophys.\ \textbf{466}, 771 (2007)] for EII agree with the experimental results to within the experimental uncertainties. Common features in the PR and EII resonance structures are identified and discussed.
\end{abstract}

\pacs{34.80.Dp, 34.80.Lx, 52.20.Fs, 95.30.Dr}


\date{\today}
\maketitle

\section{Introduction \label{sec:intro}}

Iron is the most abundant element in the cosmos heavier than silicon \cite{Asplund2009}. Features from multiply charged iron ions are present in X-ray spectra from many astrophysical sources \cite{Kahn2002a}. High-quality atomic data are required for a meaningful interpretation of the observed spectra \cite{Kallman2007a,Kallman2013}. As part of our ongoing effort to provide reliable ionization and recombination data for astrophysical applications \cite{Savin2007d,Schippers2009a,Schippers2010,Hahn2014} we here present new experimental rate coefficients for photorecombination (PR) and electron-impact ionization (EII) of Mg-like Fe$^{14+}$ ions forming Fe$^{13+}$ and Fe$^{15+}$, respectively.

For these studies, we have employed an electron-ion merged beams technique at a heavy-ion storage ring for electron-ion collision energies in the range 0--2600~eV.  Beam storage and beam cooling allow the preparation of ion beams with well characterized mass, charge, and velocity distribution. An advantage of the storage-ring technique over conventional single-pass experiments is that long-lived excited states, that often contaminate ion beams, have sufficient time to decay while the ion beam coasts in the storage ring before a cross-section measurement is started (see, e.g., \cite{Linkemann1995a,Hahn2012a}). We exploit this feature to reduce the fraction of metastable Fe$^{14+}$([Ne]\,$3s\,3p\;^3\!P$) ions in the stored ion beam.

The present recombination measurements greatly extend the energy range of our earlier experimental PR results \cite{Lukic2007a} which were limited to energies between 0 and 45~eV. To the best of our knowledge there are no previous experimental results available for EII of Fe$^{14+}$. Both PR and EII of Fe ions with this and neighboring charges are important for the abundance of charge states and the spectral line intensities in astrophysical plasmas over a large temperature range. In order to satisfy the astrophysical data needs, plasma rate coefficients are derived from the results of the present merged-beams PR and EII experiments. These experimentally derived plasma rate coefficients and the underlying cross sections can serve as benchmarks for theoretical calculations. Here, they are compared with data from the literature, in particular, with the most recent theoretical results of Altun \etal \cite{Altun2007a} for PR of Fe$^{14+}$ and of Dere \cite{Dere2007} for EII of Fe$^{14+}$.

The paper is organized as follows. The subsequent sections \ref{sec:introPR}, \ref{sec:introEII}, and \ref{sec:introCommon} introduce the specific aspects of PR and EII that are relevant for the present study.  Section \ref{sec:exp} describes the experimental setup. In \sref{sec:results} experimental results are presented, discussed and compared with literature data. Finally, a summary and conclusions are provided in \sref{sec:summary}.

\subsection{Photorecombination}\label{sec:introPR}

In the experimental energy range covered, various recombination and ionization mechanisms lead to specific features in the measured cross sections or rate coefficients \cite{Mueller2008a}. The most relevant PR processes for Fe$^{14+}$([Ne]\,$3s^2$) ions are radiative recombination (RR -- free electron capture by simultaneous photon emission), dielectronic recombination (DR -- electron capture and simultaneous excitation of one bound  electron followed by radiative stabilization) and trielectronic recombination \cite{Schnell2003b} (TR -- electron capture and simultaneous excitation of two bound electrons followed by radiative stabilization). In detail, the relevant recombination channels are:
\begin{widetext}
\begin{numcases}{\label{eq:rec}\mathrm{Fe}^{14+}(2s^2\,2p^6\,3s^2)+e^-\rightarrow}
    \mathrm{Fe}^{13+}(2s^2\,2p^6\,3s^2\,nl) + \text{photon}  ~ & RR\\
    \mathrm{Fe}^{13+}(2s^2\,2p^6\,3s\,3l'\,nl)  \rightarrow\mathrm{Fe}^{13+} + \text{photons} & $\Delta N=0$ DR ($3s\to3l'$)\label{eq:DR0}\\
    \mathrm{Fe}^{13+}(2s^2\,2p^6\,3s\,4l'\,nl) \rightarrow\mathrm{Fe}^{13+} + \text{photons} & $\Delta N=1$ DR ($3s\to4l'$)\label{eq:DR3s} \\
    \mathrm{Fe}^{13+}(2s^2\,2p^5\,3s^2\,3l'\,nl) \rightarrow\mathrm{Fe}^{13+} + \text{photons} & $\Delta N=1$ DR ($2p\to3l'$)\label{eq:DR2p} \\
    \mathrm{Fe}^{13+}(2s^\mathit{\phantom{2}}\,2p^6\,3s^2\,3l'\,nl) \rightarrow\mathrm{Fe}^{13+} + \text{photons} & $\Delta N=1$ DR ($2s\to3l'$)\label{eq:DR2s}\\
    \mathrm{Fe}^{13+}(2s^2\,2p^6\,3p^2\,nl)\rightarrow\mathrm{Fe}^{13+} + \text{photons} & $\Delta N=0$ TR ($3s^2\to3p^2$)\label{eq:TR}.
\end{numcases}
\end{widetext}

Here and below, $n'$ and $l'$ denote the principal quantum number and orbital angular momentum quantum number, respectively, of the electron excited in the parent ion upon recombination (core excited electron), and $n l$ are the analogous quantum numbers of the captured (Rydberg) electron. For DR a distinction is made whether the core electron is excited within an atomic shell ($\Delta N=0$ DR) or whether it is excited to the next higher shell ($\Delta N=1$ DR). Our previous experiment \cite{Lukic2007a} addressed only $\Delta N=0$ DR and TR. Here we additionally consider $\Delta N=1$ DR involving the excitation of a $2s$, $2p$, or $3s$ core electron. These DR channels are particularly important in collisionally ionized plasmas such as the solar corona \cite{Savin2007d}. In principle DR involving higher excitations ($\Delta N\ge 2$ DR) is also possible, however, we find that the corresponding contributions to the measured rate coefficients are negligible. Likewise, we do not find any signatures of $\Delta N\ge 1$ TR.

Except for the results of our previous study of $\Delta N=0$ DR of Fe$^{14+}$ \cite{Lukic2007a} there are no further experimental PR rate coefficients available for Mg-like ions. Theoretical calculations for this isoelectronic sequence of ions have been carried out by Altun \etal \cite{Altun2007a} for ions with nuclear charges $13\leq Z \leq54$. That publication also summarizes earlier theoretical work on DR of Mg-like ions.

\subsection{Electron-impact ionization}\label{sec:introEII}

In the present experiment, EII of Fe$^{14+}$([Ne]\,$3s^2$) involves \cite{Mueller2008a} direct ionization (DI), excitation autoionization (EA), resonant excitation double autoionization (REDA), and resonant excitation auto double ionization (READI):
\begin{widetext}
\begin{numcases}{\label{eq:ion}\mathrm{Fe}^{14+}(2s^2\,2p^6\,3s^2)+e^-  \rightarrow }
    \mathrm{Fe}^{15+}(2s^2\,2p^6\,3s) + 2e^- &DI ($3s$)\label{eq:DI3s}\\
    \mathrm{Fe}^{15+}(2s^2\,2p^5\,3s^2) + 2e^- & DI  ($2p$) \label{eq:DI2p}\\
    \mathrm{Fe}^{15+}(2s^\mathit{\phantom{2}}\,2p^6\,3s^2) + 2e^- &   DI  ($2s$)\label{eq:DI2s}\\
    \mathrm{Fe}^{14+}(2s^2\,2p^5\,3s^2\,n'l')+e^-\rightarrow\mathrm{Fe}^{15+}(2s^2\,2p^6\,3s) +2e^-  & EA ($2p\to n'l'$)~~\label{eq:EA2p}\\
    \mathrm{Fe}^{14+}(2s^\mathit{\phantom{2}}\,2p^6\,3s^2\,n'l')+e^-\rightarrow\mathrm{Fe}^{15+}(2s^2\,2p^6\,3s) +2e^- &EA ($2s\to n'l'$)~~\label{eq:EA2s}\\
    \mathrm{Fe}^{13+}(2s^2\,2p^5\,3s^2\,n'l'\,nl) \rightarrow \mathrm{Fe}^{14+}(2s^2\,2p^5\,3s^2\,n''l'') + e^- & \nonumber\\
    ~\hspace*{4.35cm}\rightarrow \mathrm{Fe}^{15+}(2s^2\,2p^6\,3s) + 2e^- &REDA\label{eq:REDA}\\
    \mathrm{Fe}^{13+}(2s^2\,2p^5\,3s^2\,n'l'\,nl) \rightarrow \mathrm{Fe}^{15+}(2s^2\,2p^6\,3s) +2e^- &READI\label{eq:READI}.
\end{numcases}
\end{widetext}

The experimental energy range comprises the thresholds for DI of $3s$, $2p$, and $2s$ electrons at about 454, 1179, and 1309~eV, respectively, as calculated with Cowan's atomic structure code \cite{Cowan1981}. According to the same calculations, EA associated with the excitation of a $2s$ or $2p$ electron is  possible at energies above 731~eV and 872~eV, respectively. Ionization channels involving K-shell electrons open up only at the $1s\to3p$ excitation energy of 7162~eV which is far beyond the highest electron-ion collision energy accessed in the present experiment.

EII of Mg-like ions has received somewhat more attention in the literature \cite{Dere2007} than has recombination. There are no experimental cross sections available for EII of  Fe$^{14+}$ but there are for some lower charged Mg-like ions such as Al$^+$ \cite{Montague1983,Belic1987a,Hayton1994}, Si$^{2+}$ \cite{Djuric1993}, S$^{4+}$ \cite{Howald1986a}, Cl$^{5+}$ \cite{Howald1986a}, and Ar$^{6+}$ \cite{Howald1986a,Zhang2002,Becker2012}. All these experiments employed single-path setups with, in some cases, rather large fractions of ions in long-lived $3s\,3p\;^3\!P$ levels in the primary ion beams. The fact that these fractions were not known compromised the derivation of absolute cross sections from these measurements to some extent.

Theoretical EII cross sections for Mg-like ions, including the work of Mitnik \etal \cite{Mitnik1998a} for Fe$^{14+}$, have been reviewed by Dere \cite{Dere2007} who also performed theoretical calculations for $12\leq Z \leq 30$ and derived recommended rate coefficients for use in astrophysical modeling. Since then new theoretical cross sections have been published for Mg-like Al$^+$ \cite{Ludlow2009}, Si$^{2+}$ \cite{Colgan2008a}, Ar$^{6+}$ \cite{Loch2007b}.

\subsection{Common features in PR and EII cross sections}\label{sec:introCommon}

Another aspect of the present study is the search for common recombination and ionization features as has already been pursued by Linkemann~\etal~\cite{Linkemann1995c} for Na-like Fe$^{15+}$. Inspection of \erefs{eq:rec}{eq:ion}{ and } reveals that the same intermediate states may be involved in recombination and ionization. For example Fe$^{13+}$($2p^5\,3s^2\,3p\,nl$) states are transiently populated in $2p\to3l'$ DR [\eref{eq:DR2p}] as well as in ionization via REDA or READI [\erefs{eq:REDA}{eq:READI}{ and } with $n'=3$]. DR, REDA, and READI are all resonant processes. Corresponding resonance structures may thus be observed in both the total cross sections for PR and EII with the relative resonance strength depending on the branching ratios for radiative and autoionizing decay of the intermediate resonance state. Another correspondence exists between EA thresholds [\erefs{eq:EA2p}{eq:EA2s}{ and } with $n'=3$] and DR series limits [\erefs{eq:DR2p}{eq:DR2s}{ and } with $n\to\infty$]. The present experimental approach of simultaneous PR and EII measurements directly allows one to identify such related features in PR and EII cross sections.

\section{Experiment\label{sec:exp}}

The PR and EII measurements were performed at the heavy-ion storage ring TSR \cite{Grieser2012} of the Max-Planck-Institut f\"{u}r Kernphysik (MPIK) in Heidelberg, Germany.
The experimental procedures were similar to those applied in previous measurements \cite{Lukic2007a,Schippers2001c,Lestinsky2012,Hahn2010} (and references therein). Here we focus on the details that are specific to the present measurements.

The MPIK tandem and linear accelerators were used to accelerate iron ions to a final energy of about 160~MeV. Along the way, multiply charged iron ions were produced by electron stripping in thin carbon foils. Ions with the desired mass-to-charge ratio were selected by passing the beam through a dipole magnet with the magnetic field strength and the widths of beam collimating slits adjusted such that only $^{57}$Fe$^{14+}$ ions were injected into the storage ring. The choice of this particular isotope with a natural abundance of only 2.1\% \cite{Boehlke2005a} is motivated  below (\sref{sec:meta}).

In two of the straight sections of the TSR the ion beam was merged with two separate electron beams. The two electron beam devices are referred to as \lq Cooler\rq\ and \lq Target\rq. Both can be used as an electron cooler \cite{Poth1990} and as a target for electron-ion collision experiments. During the present measurements the stored ions were continuously cooled by the Target electron beam. To this end the Target electron energy was set to the cooling energy $E_\mathrm{cool} = 1527.8$~eV where electrons and $^{57}$Fe$^{14+}$ ions moved with the same average velocity. The storage lifetime of the cooled beam was about 1~min.

The Cooler was chosen as a target for the PR and EII measurements because it provides a higher electron density and thus larger signal count rates than the Target.
The first TSR dipole magnet behind the Cooler was used to separate Fe$^{13+}$ recombination and Fe$^{15+}$ ionization products from the stored Fe$^{14+}$ ion beam. The different reaction products were counted by two appropriately placed single-particle detectors \cite{Rinn1982} with nearly 100\% efficiency.

After injection of ion pulses into the TSR and subsequent initial beam cooling for typically 2~s, measurements over a range of electron-ion collision energies were performed by stepping the electron acceleration voltage $U$ at the Cooler through a preselected set of up to 1000 values $U_{k=1..k_\mathrm{max}}$. Between any two measurement voltages $U_k$ and $U_{k+1}$ the acceleration voltage was set to a suitably chosen reference voltage $U_\mathrm{ref}$. The recombination and ionisation signals associated to $U_\mathrm{ref}$ were used  for background determination \cite{Novotny2012,Hahn2010}. There was an 11~ms waiting time interval after each change of $U$ to allow the power supplies to settle to their new values. Data taking took place during the subsequent 10~ms, before $U$ was changed to the next value. The entire injection-cooling-measurement sequence was repeated until suitable numbers of product counts had been accumulated at each measurement energy. Multiple measurement scan ranges with $\sim50\%$ overlap were used to cover electron-ion collision energies between 0 and 2600~eV.

Usually absolute merged-beams rate coefficients for PR and EII are readily derived from the measured recombination and ionization count rates  by normalization on ion current and on Cooler electron current \cite{Schippers2001c,Hahn2010}. Because of the use of the less abundant $^{57}$Fe isotope the stored ion current was rather low. It typically amounted to 1--3~$\mu$A after the initial cooling  and from then on decreased almost exponentially with a time constant of about 1~min. Such low currents are below the sensitivity threshold of the beam-current transformer that is usually used for the ion current measurement. In this situation a signal that was proportional to the stored ion current was derived from the beam profile monitor (BPM) which is routinely used for non-destructive measurements of horizontal and vertical ion-beam profiles \cite{Hochadel1994a}. The working principle of the BPM is based on residual-gas ionization by the stored ions. Thus, the BPM count rate is proportional to ion current and residual-gas density and, under stable vacuum conditions, relative rate coefficients can be obtained by normalisation to the BPM signal. The present relative rate coefficients were put on an absolute scale by normalisation to the result of our previous low-energy recombination measurement \cite{Lukic2007a}. It should be noted that the same normalization constant applies to both the PR and the EII rate coefficient. The uncertainty of both rate coefficients is therefore given by the 26\%  uncertainty of our previous measurement \cite{Lukic2007a}. Here and throughout all uncertainties are quoted at a level comparable to a 90\% statistical accuracy.

The systematic uncertainty of the experimental energy scale depends on the accuracy of the power supplies that provide the acceleration potential for the Cooler electrons.  A comprehensive discussion of all sources of uncertainty has been given by Kilgus~\etal~\cite{Kilgus1992}. The largest uncertainties originate from beam-alignment and space-charge effects. The relative error of the collision energy scale increases with the collision energy, being smallest close to 0~eV \cite{Lestinsky2008a}. For the rather high electron-ion collision energies of the present experiment we assume a systematic relative error of up to 0.5\%. This uncertainty may be reduced if known spectral features can be used for the calibration of the collision energy scale (see below).

\subsection{Metastable ions}\label{sec:meta}

As already discussed in our earlier work on PR of Fe$^{14+}$~\cite{Lukic2007a} Mg-like ions are known to have long-lived $3s\,3p\;^3\!P_J$ levels ($J=0,1,2$) which might have been present in the ion beam. The lifetimes of the $^3\!P_1$ and $^3\!P_2$ levels are $\sim2.6\times10^{-8}$~s  and $\sim2.5\times10^{-2}$~s, respectively \cite{Aggarwal2007a}, i.e., much shorter than the initial 2~s cooling time. In contrast, the lifetime of the $^3\!P_0$ level is considerably longer (practically infinitely long) because the  $^3\!P_0\to {^1\!S_0}$ one-photon decay is forbidden. For nuclei with nonzero magnetic moment, however, hyperfine quenching shortens the $^3\!P_0$ lifetime. A recent calculation yields a value of $\sim$21~s for the hyperfine induced lifetime of the $^{57}$Fe$^{14+}$($3s\,3p\;^3\!P_0$) level \cite{Andersson2010}.

In fact, one motivation for the present recombination measurements was to experimentally measure this lifetime using the same approach as already successfully applied for the measurements of hyperfine induced lifetimes of the $2s\,2p\;^3\!P_0$ levels in Be-like $^{47}$Ti$^{18+}$~\cite{Schippers2007a} and $^{33}$S$^{12+}$~\cite{Schippers2012}. This approach requires the identification of DR resonances of $^3\!P_0$ parent ions by comparing recombination spectra of $^{56}$Fe$^{14+}$ and $^{57}$Fe$^{14+}$ ions. If such resonances could be observed for $^{56}$Fe$^{14+}$ their presence in the $^{57}$Fe$^{14+}$ DR spectrum would be exponentially suppressed with time elapsing after injection into the storage ring, i.e., with the $3s\,3p\;^3\!P_0$ level decaying via the hyperfine-induced transition. However, no significant differences between the measured DR resonance structure of $^{56}$Fe$^{14+}$ and $^{57}$Fe$^{14+}$ were found. This is attributed to the fact that DR resonances of $^3\!P_0$ parent ions have been predicted to be only weak and to be blended with the rich DR resonance structure of $^1\!S_0$ ground-state parent ions \cite{Lukic2007a}.

\begin{figure}[t]
\includegraphics[width=\figurewidth]{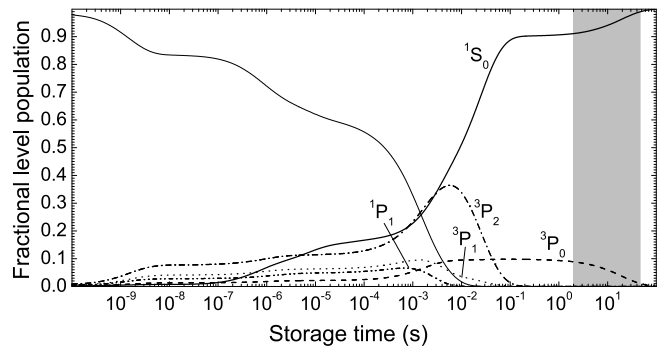}
\caption{\label{fig:pop} Calculated time dependent fractional populations of the $^{57}$Fe$^{14+}$($3s^2\;^1\!S_0$) ground level (thick solid line) and of the excited levels $3s\,3p\;^3\!P_0$, $^3\!P_1$, $^3\!P_2$ and $^1\!P_1$ (dashed line, dotted line, dash-dotted line, and dash-dot-dotted line, respectively). The thin solid line is the sum of the populations of the 136 higher excited levels that were also considered in the calculation. The shaded area marks the time interval that was typically used for data taking.}
\end{figure}

In our previous work \cite{Lukic2007a} we estimated that $6\%\pm6\%$ of the stored ions had been in the $3s\,3p\;^3\!P_0$ level and the remaining fraction in the $3s^2\;^1\!S$ ground state. The 6\% uncertainty is included in the total 26\% error budget. Since we have normalized the present PR and EII rate coefficients to our previous results the same uncertainty applies here. It should be pointed out, that the normalization could be carried out unambiguously, since there was no visible difference --- apart from differences due to different experimental energy spreads --- between the previously measured DR resonance structure of $^{56}$Fe$^{14+}$ and the present one for $^{57}$Fe$^{14+}$.

The 6\% estimate for the $3s\,3p\;^3\!P_0$ fraction was based on assuming initial statistical populations of the $^3\!P_J$ levels. Here we refine the estimate by taking radiative transitions between levels into account. The population dynamics is obtained by solving a set of rate equations \cite{Lestinsky2012} with radiative transitions linking higher excited levels to lower levels. Here, we used the dipole (E1) transition rates from \cite{Aggarwal2007a} between the 141 lowest levels from the $3l^2$, $3l\,3l'$, and $3l\,4l'$ configurations and the hyperfine induced $3s\,3p\;^3\!P_0\to3s^2\;^1\!S_0$ transition rate from \cite{Andersson2010}. Furthermore, we assumed that the foil-stripping process produced a Boltzmann distribution of initial populations \cite{Lestinsky2012} of all 141 levels. The resulting population curves are shown in figure \ref{fig:pop}.

\begin{figure*}
\includegraphics[width=0.9\textwidth]{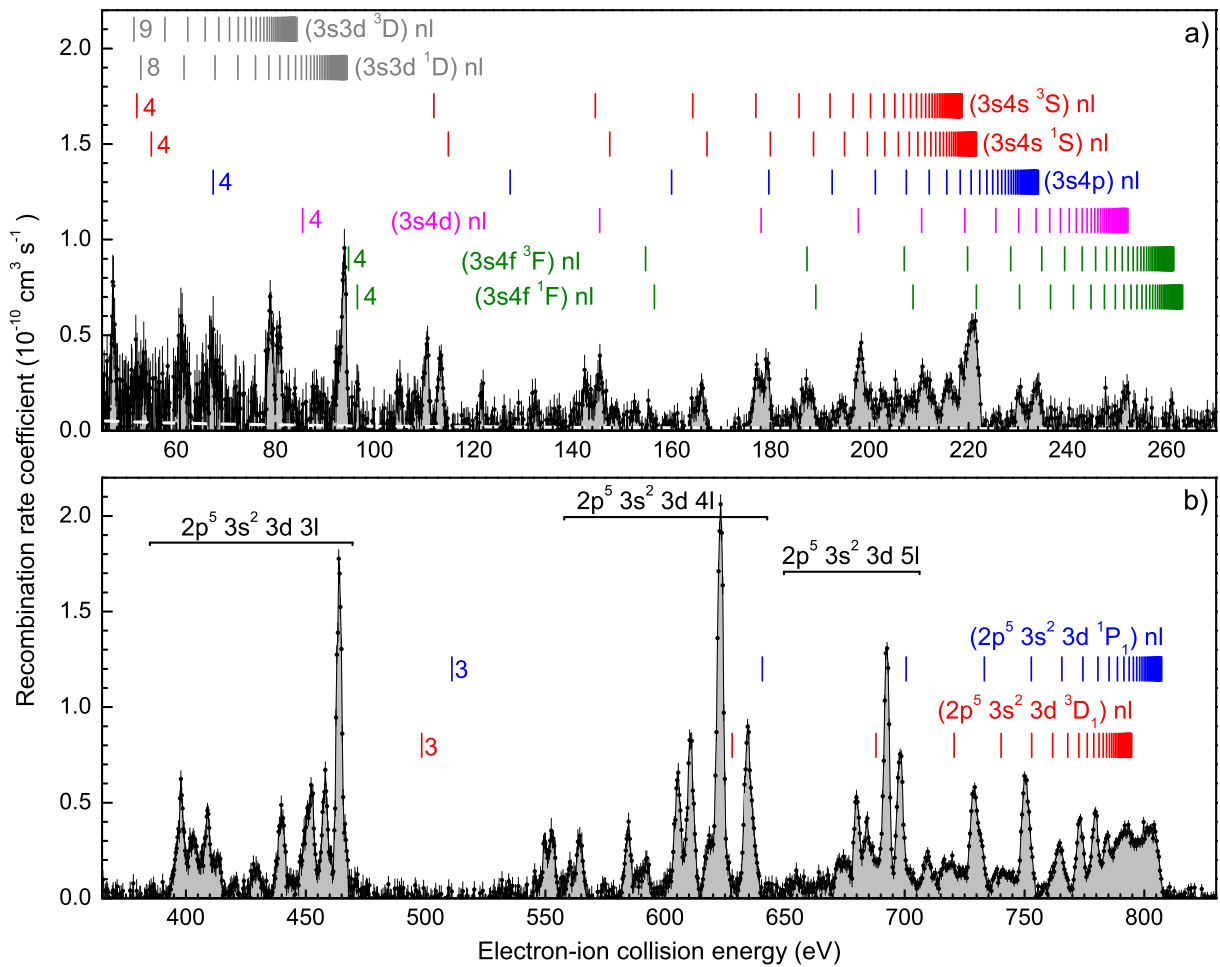}
\caption{\label{fig:rec} (color online) Measured merged-beams rate coefficient for PR of Fe$^{14+}$ in the energy ranges of the (a) $3\to4$ $\Delta N=1$ DR resonances [\eref{eq:DR3s}] and (b) $2\to3$ $\Delta N=1$ DR resonances [\erefs{eq:DR2p}{eq:DR2s}{ and }]. The vertical bars denote calculated resonance positions
[\eref{eq:Ryd}]. Numbers at the leftmost vertical bars denote the value of the lowest Rydberg principal quantum number $n$ considered. Some of the resonances in
panel (a) may be associated with $\Delta N=0$ DR $3s\to3d$ core excitations. The dashed white line in panel (a) is the theoretical rate coefficient for RR of Fe$^{14+}$ calculated by using a semi-classical hydrogenic formula with field-ionization cutoff beyond $n_\mathrm{max} = 53$ \cite{Schippers2001c}. The RR rate coefficient has not been plotted in panel (b) where it is practically zero on the displayed rate-coefficient scale.}
\end{figure*}

As expected, most excited levels have decayed within a storage time of 0.1~s. At later times only the  $3s\,3p\;^3\!P_0$ and the $3s^2\;^1\!S_0$ levels are populated. Up to about 20~s the corresponding fractions amount to about 10\% and 90\%, respectively. These fractions have been found to be largely insensitive to variations of the initial populations. At even later times the hyperfine induced transition depopulates the $3s\,3p\;^3\!P_0$ level. Its population is practically zero after 60~s, i.e., after one storage lifetime. Since the 0--10\% population of the $3s\,3p\;^3\!P_0$ level is within the uncertainty of the assumed value of $6\%\pm6\%$ these findings do not compromise our normalization procedure.

Clearly, an initial cooling time of about 60~s would have facilitated measurements with pure $^{57}$Fe$^{14+}$($3s^2\;^1\!S_0$) beams. Unfortunately, the relatively short storage lifetime of the ion beam prevented us from delaying data taking for a sufficiently long time as has been done in~\cite{Hahn2012a}.

\section{Results\label{sec:results}}

\subsection{Recombination\label{sec:rec}}

The Fe$^{14+}$ merged-beams recombination rate coefficient was obtained for energies between 0 and 2600~eV.  The energy range 0--42~eV, which comprises the most important $\Delta N =0$ DR resonances [\eref{eq:DR0}], has already been studied in detail in our previous work \cite{Lukic2007a}. Therefore, we here focus on higher electron-ion collision energies where DR is associated with $\Delta N=1$ core excitations [\erefs{eq:DR3s}{eq:DR2s}{--}]. Our previous measurements were performed with a colder but less dense electron beam. For the present experiment a high-density electron beam was used which allowed us to investigate smaller cross sections at the expense of energy resolution. The measured merged-beams recombination rate coefficient is displayed in figure~\ref{fig:rec}. The two featured energy ranges comprise all measured $3\to4$ and $2\to3$ $\Delta N=1$ DR resonances [\erefs{eq:DR3s}{eq:DR2s}{--}]. Within the statistical experimental uncertainties no resonances were detected in the energy ranges 290--365~eV and 830--2600~eV. In order to gain some insight into the measured resonance structures we estimated DR resonance positions using the Rydberg formula for resonance energies $E_n$, i.e.,
\begin{equation}\label{eq:Ryd}
E_n = E_\mathrm{ex}-\mathcal{R}\frac{q^2}{n^2}
\end{equation}
where $q=14$ is the charge of the primary Fe$^{14+}$ ion  and $\mathcal{R} = 13.6057$~eV. Core excitation energies $E_\mathrm{ex}$ from the literature and from atomic-structure calculations are listed in \tref{tab:Eexci}.

\begin{table}
\caption{\label{tab:Eexci} Energies for selected one-electron excitations of Fe$^{14+}$([Ne]\,$3s^2\;^1\!S_0$) ions. The M-shell excitation energies in the second column are from \cite{Landi2011}. The L-shell exciation energies in the forth column have been calculated using Cowan's atomic structure code \cite{Cowan1981} in the single-configuration approximation.}
\begin{ruledtabular}
\begin{tabular}{cd|cd}
\multicolumn{2}{c|}{$3\to3$ and $3\to4$ excitations} & \multicolumn{2}{c}{$2\to3$ excitations} \\
	Term & \multicolumn{1}{c|}{$E_\mathrm{ex}$ (eV)} & Level & \multicolumn{1}{c}{$E_\mathrm{ex}$ (eV)} \\
    \hline
	$3s\,3d\;^3\!D$ &  84.35 & $2p^5\,3s^2\,3p\;^3\!S_1$ & 730.91 \\
	$3s\,3d\;^1\!D$ &  94.49 & $2p^5\,3s^2\,3p\;^3\!D_2$ & 733.17 \\
	$3s\,4s\;^3\!S$ & 218.70 & $2p^5\,3s^2\,3p\;^3\!D_3$ & 734.64 \\
	$3s\,4s\;^1\!S$ & 221.61 & $2p^5\,3s^2\,3p\;^1\!P_1$ & 735.60 \\
	$3s\,4p\;^3\!P$ & 233.90 & $2p^5\,3s^2\,3p\;^3\!P_2$ & 737.07 \\
	$3s\,4p\;^1\!P$ & 234.31 & $2p^5\,3s^2\,3p\;^3\!P_0$ & 742.57 \\
	$3s\,4d\;^3\!D$ & 252.01 & $2p^5\,3s^2\,3p\;^3\!D_1$ & 745.27 \\
	$3s\,4d\;^1\!D$ & 252.34 & $2p^5\,3s^2\,3p\;^3\!P_1$ & 748.14 \\
	$3s\,4f\;^3\!F$ & 261.45 & $2p^5\,3s^2\,3p\;^1\!D_2$ & 748.46 \\
	$3s\,4f\;^1\!F$ & 263.23 & $2p^5\,3s^2\,3p\;^1\!S_0$ & 760.87 \\
                  &        & $2p^5\,3s^2\,3d\;^3\!P_0$ & 783.79 \\
                  &        & $2p^5\,3s^2\,3d\;^3\!P_1$ & 784.68 \\
                  &        & $2p^5\,3s^2\,3d\;^3\!F_4$ & 786.12 \\
                  &        & $2p^5\,3s^2\,3d\;^3\!P_2$ & 786.26 \\
                  &        & $2p^5\,3s^2\,3d\;^3\!F_3$ & 786.73 \\
                  &        & $2p^5\,3s^2\,3d\;^1\!D_2$ & 788.19 \\
                  &        & $2p^5\,3s^2\,3d\;^3\!D_3$ & 789.14 \\
                  &        & $2p^5\,3s^2\,3d\;^3\!D_1$ & 793.53\footnote{Experimental value: 794.95(10) eV \cite{Beiersdorfer2013}}\\
                  &        & $2p^5\,3s^2\,3d\;^3\!F_2$ & 799.19 \\
                  &        & $2p^5\,3s^2\,3d\;^3\!D_2$ & 799.97 \\
                  &        & $2p^5\,3s^2\,3d\;^1\!F_3$ & 800.42 \\
                  &        & $2p^5\,3s^2\,3d\;^1\!P_1$ & 806.83\footnote{Experimental value: 807.29(05) eV \cite{Beiersdorfer2013}}\\
                  &        & $2s\,2p^6\,3s^2\,3p\;^3\!P_0$ & 871.48 \\
                  &        & $2s\,2p^6\,3s^2\,3p\;^3\!P_1$ & 871.91 \\
                  &        & $2s\,2p^6\,3s^2\,3p\;^3\!P_2$ & 873.99 \\
                  &        & $2s\,2p^6\,3s^2\,3p\;^1\!P_1$ & 875.74 \\
                  &        & $2s\,2p^6\,3s^2\,3d\;^3\!D_1$ & 924.87 \\
                  &        & $2s\,2p^6\,3s^2\,3d\;^3\!D_2$ & 925.03 \\
                  &        & $2s\,2p^6\,3s^2\,3d\;^3\!D_3$ & 925.31 \\
                  &        & $2s\,2p^6\,3s^2\,3d\;^1\!D_2$ & 929.37 \\
\end{tabular}
\end{ruledtabular}
\end{table}

Very few of the many possible excitation channels can be identified in the experimental data. In \fref{fig:rec}(a) series limits can be discerned for the $3s\,4s\,nl$ and $3s\,4p\,nl$ Rydberg series of DR resonances at about 221 and 234~eV, respectively. Another feature at about 251~eV is suggestive of the $3s\,4d\,nl$ series limit. There is no clear indication for the $3s\,4f\,nl$ Rydberg series limit. The experimental resolving power is not sufficient for discrimination between all terms of the $3l'\,4l$ configurations. An unambiguous assignment of the measured resonances that appear below these series limits is difficult because of the rather large statistical uncertainties of the measured merged-beams rate coefficient in the energy range of \fref{fig:rec}(a). The strongest resonances align best with the estimated $3s\,4d\,nl$ resonance positions.  For low $n$, the $nl$ Rydberg electron interacts strongly with the core which leads to a large energy splitting of the resonances within the Rydberg manifold and makes the Rydberg formula [\eref{eq:Ryd}] inadequate. Therefore, the features at 80 and 93~eV may be assigned as $3s\,4l'\,4l$ resonances. On the other hand these features also coincide with the $3s\,3d\,nl$ $\Delta N=0$ DR series limits. However, we consider their contribution to be insignificant as no resonances from this Rydberg series, which extends down to energies as low as about 10~eV, could be identified in our previous work \cite{Lukic2007a}. For the derivation of the plasma rate coefficient (see \fref{sec:plasmaDR}) we treat all resonances from \fref{fig:rec}(a) as $3\to4$ $\Delta N=1$ DR resonances [\eref{eq:DR3s}].

Figure \ref{fig:rec}(b) displays the measured merged-beams recombination rate coefficient in the energy range of the $2\to3$ $\Delta N=1$ DR resonances [\erefs{eq:DR2p}{eq:DR2s}{ and }]. Two Rydberg series limits are discernible at about 795 and 807~eV associated with $2p\to3d$ core excitations to the $2p^5\,3s^2\,3d\;^3\!D_1$ and $2p^5\,3s^2\,3d\;^1\!P_1$ levels, respectively (\tref{tab:Eexci}).
The radiative rates from these excited levels to the $3s^2\;^1\!S_0$ ground states are the largest of all $2\to3$ excitations to levels with total angular momentum $J=1$ \cite{Gu2006}. The corresponding spectral lines have been observed in an optical measurement at an electron-beam ion trap \cite{Beiersdorfer2013} which provided experimental values for the $2p^5\,3s^2\,3d\;^3\!D_1$ and $2p^5\,3s^2\,3d\;^1\!P_1$ excitation energies with rather low uncertainties (\tref{tab:Eexci}). We have used these values for a more accurate calibration of the experimental energy scale of \fref{fig:rec}. The calibration consists in a multiplication of the nominal collision-energy scale by a factor 1.0045. This factor, which is within the uncertainty of the nominal energy scale (\sref{sec:exp}), was chosen such that calculated [\eref{eq:Ryd} with $E_\mathrm{ex}=807.29$~eV] and measured  $(2p^5\,3s^2\,3d\;^1\!P_1)\,nl$ Rydberg resonance features in the 750--810~eV energy range line up as well as achievable. This factor was also applied to the electron-ion collision energy scale of the measured EII cross section below.

The resonance structures at lower energies, i.e., in the range of about 380--750~eV, do not exhibit any regularity. The given assignments in \fref{fig:rec}(b) are meant to only serve as a coarse orientation. In particular, DR associated with $2p\to3p$ and $2s\to3l'$ core excitations may play some role, although corresponding series limits cannot be identified in the measured data.

\subsubsection{Plasma rate coefficient for DR of Fe$^{14+}$\label{sec:plasmaDR}}

 \begin{figure}[t]
\includegraphics[width=\figurewidth]{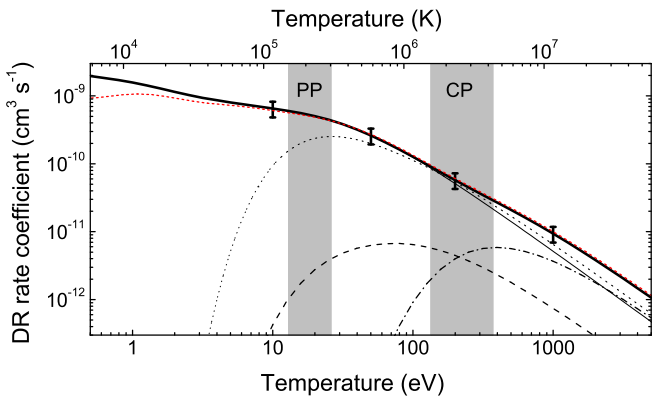}
\caption{\label{fig:plasmaDR} (color online) Rate coefficients for DR of Fe$^{14+}$($3s^2\;^1\!S_0$) ions in a plasma. The thick full line is our present experimentally derived rate coefficient. The error bars denote its $\pm26\%$ uncertainty. Our previous result \cite{Lukic2007a} that only comprised $\Delta N=0$ DR [\eref{eq:DR0}] is shown as thin full line. The present additional contributions by $3\to4$ $\Delta N=1$ DR [\eref{eq:DR3s}] and by $2\to3$ $\Delta N=1$ DR [\erefs{eq:DR2p}{eq:DR2s}{ and }] are shown as dashed and dash-dotted curves, respectively. The thin dotted curve is the recommended rate coefficient from  Arnaud and Raymond \cite{Arnaud1992}. The most recent theoretical result of Altun \etal~\cite{Altun2007a} is shown as (red) short-dashed line. The temperature ranges where Fe$^{14+}$ is expected to form in photoionized plasmas (PP) \cite{Kallman2010} and collisionally ionized  plasmas (CP) \cite{Bryans2009a} are indicated as gray shaded areas.}
\end{figure}

We have derived DR plasma rate coefficients from the $3\to4$ and $2\to3$ $\Delta N=1$ data in \fref{fig:rec} by following the procedures described already earlier \cite{Schippers2001c}.  In particular, we have subtracted a smooth theoretical rate coefficient for RR [dashed line in \fref{fig:rec}(a)] from the measured merged beams rate coefficient. The remaining merged-beams DR rate coefficient was converted into a cross section and then convoluted with a Maxwellian electron energy distribution. The resulting $3\to4$ and $2\to3$ DR rate coefficients in a plasma are plotted in \fref{fig:plasmaDR} as dashed and dash-dotted lines, respectively. These have been added to our previous plasma rate coefficient for $\Delta N=0$ DR (thin full line in \fref{fig:plasmaDR}) to yield our new experimentally derived DR rate coefficient (thick full line in \fref{fig:plasmaDR}).

It should be noted that the effect of field-ionization in the storage-ring bending magnets which suppresses recombination into high-$n$ Rydberg states \cite{Schippers2001c} and which has been considered in our previous work on $\Delta N=0$ DR \cite{Lukic2007a} does not significantly affect $\Delta N=1$ DR \cite{Savin2002c}. Under the present experimental conditions, field ionization occurs for $n>n_\mathrm{max}=53$. Within the present work, this cutoff has only been considered for the calculation of the theoretical merged-beams RR rate coefficient in \fref{fig:rec}(a).

For convenient use of our result in plasma modelling codes we provide a simple parametrization where we have fitted the function
\begin{equation} \label{eq:DRfit}
\alpha_\mathrm{DR}(T) = T^{-3/2} \sum_{i=1}^{8} c_i~\mathrm{exp}(-E_i/T)
\end{equation}
to our experimentally derived DR rate coefficient.  The fit parameters $c_i$ and $E_i$ are listed in Tab. \ref{tab:DRfit}.
For temperatures between 70 and 600~K the fit deviates from the experimentally derived curve by less than 1.6\%. Between 600~K and $10^8$~K the deviation is less than 1\%. This latter temperature range comprises the temperature ranges where Fe$^{14+}$ is expected to exist in photoionized and collisionally ionized plasmas (shaded areas in \fref{fig:plasmaDR}). The systematical uncertainty of the experimentally derived DR rate coefficient in a plasma amounts to 26\% at a 90\% confidence limit, coming directly from  the uncertainty of the experimental merged-beams recombination rate coefficient (\sref{sec:exp}).

\begin{table}[t]
\caption{\label{tab:DRfit}Parameters for the parametrization [\eref{eq:DRfit}] of the experimentally derived rate coefficient for DR of Fe$^{14+}$($3s^2\;^1\!S_0$) in a plasma.
The numbers in the square brackets denote powers of 10. The parameters are valid for plasma temperatures between 70 and $10^8$~K.}
\begin{ruledtabular}
\begin{tabular}{cdd}
   $i$ & \multicolumn{1}{r}{$c_i$ (cm$^3$\,s$^{-1}$\,K$^{3/2}$)} & \multicolumn{1}{r}{$E_i$ (K)} \\
  \hline
  1 & 1.431[-4] &	9.5222[1]   \\
  2 & 3.679[-4] &	2.3378[3]   \\
  3 & 1.756[-3]	&   8.6290[3]   \\
  4 & 5.667[-3]	&   2.4486[4]   \\
  5 & 1.657[-2]	&   8.6968[4]   \\
  6 & 4.937[-2]	&   1.9226[5]   \\
  7 & 1.433[-1]	&   5.1245[5]   \\
  8 & 2.832[-1]	&   6.5454[6]   \\
\end{tabular}
\end{ruledtabular}
\end{table}

The present $\Delta N=1$ results do not affect the Fe$^{14+}$ DR rate coefficient in the temperature range where Fe$^{14+}$ is expected to exist in a photoionized plasma (PP). However, $\Delta N=1$ DR is important in the temperature range relevant to collisionally ionized plasmas (CP) where it contributes between 10\% and 25\% of the total DR rate coefficient. In this temperature range the Fe$^{14+}$ DR rate coefficient from the widely used compilation of Arnaud and Raymond \cite{Arnaud1992} also agrees with the experiment within the experimental uncertainties. This latter result is based on theoretical calculations that were geared towards $\Delta N=1$ DR but did not accurately describe low-energy $\Delta N=0$ DR. Consequently, in the PP temperature range, the DR rate coefficient of Arnaud and Raymond is up to a factor of 4 lower than our experimentally derived rate coefficient. In contrast, the result from the latest state-of-the-art calculation by Altun \etal \cite{Altun2007a} agrees with the present experimentally derived DR rate coefficient within the experimental uncertainties over a wide range of plasma temperatures, including both the relevant temperature ranges for PP and CP.  Unlike the other theories, it reproduces the experimental rate coefficient also above 1000~K.

\subsection{Ionization\label{sec:ion}}

\begin{figure*}
\includegraphics[width=0.9\textwidth]{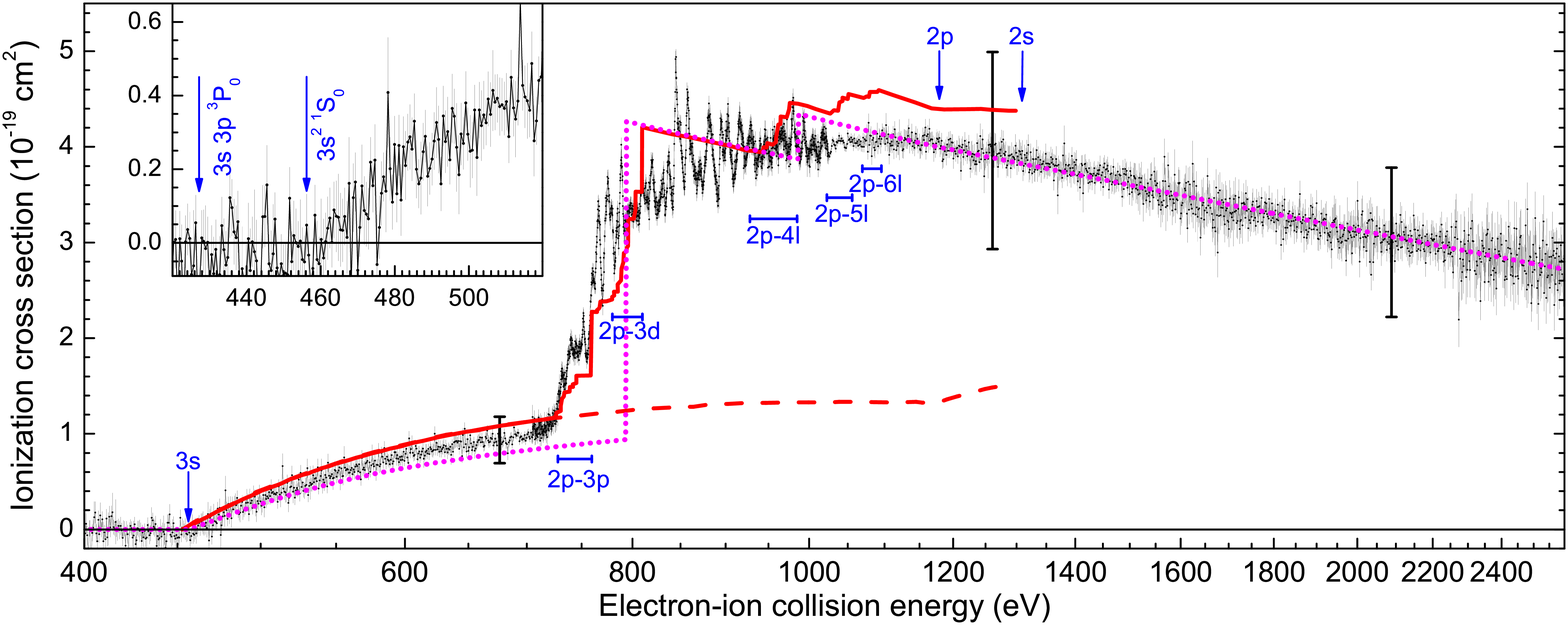}
\caption{\label{fig:EII}  (color online) Cross sections for EII of Fe$^{14+}$ forming Fe$^{15+}$. The electron-ion collision energy scale is logarithmic. The thin (black) solid line is the present experimental result. The capped error bars represent the $\pm26\%$ systematic experimental uncertainty. The dotted (magenta) line is the theoretical result of Dere \cite{*[{}] [{; the EII cross section as a function of electron-ion collision energy was retrieved from the  \textsc{chianti} data base via \url{http://www.chiantidatabase.org/}.}] Dere2007}. The thick (red) solid line is the theoretical result of Mitnik~\etal \cite{Mitnik1998a}. Their theoretical DI cross section is shown separately as a (red) dashed line. Vertical arrows mark DI ionization thresholds. Energy ranges \cite{Mitnik1998a} for the various $2p\to n'l'$ EA thresholds [\eref{eq:EA2p}] are also indicated. The inset enlarges the energy range around the threshold for DI of a $3s$ valence electron. The threshold energies in the inset were taken from \cite{Kramida2013}.}
\end{figure*}

Our experimental cross section for EII of Fe$^{14+}$($3s^2\;^1\!S_0$) ions is displayed in \fref{fig:EII} as well as the results from the distorted wave calculations for EII of Fe$^{14+}$ by Dere \cite{Dere2007} and by  Mitnik~\etal~\cite{Mitnik1998a}. The latter extends only up to the $2s$ DI threshold at about 1309~eV (\sref{sec:intro}). Within the $\pm26\%$ experimental uncertainty both of these theoretical cross sections agree with the measured cross section, with the exception of the Dere cross section in the 740--810~eV energy range. Above 1200~eV we find good agreement between the cross section of Dere and the experimental results. There are strong similarities between the overall shapes of the three cross section curves, but there are also distinct differences which are discussed in more detail below.

Within the statistical uncertainties, the measured cross section is zero below the threshold for DI of a $3s$ electron from the $3s^2$ ground configuration. In particular, there are no signs of ionization of metastable Fe$^{14+}$($3s\,3p\;^3\!P_0$) ions which could be expected below the Fe$^{14+}$($3s^2\;^1\!S_0$) ground-level ionization threshold. This finding is consistent with our assumption of a $6\%\pm6\%$ metastable fraction in the parent ion beam (\sref{sec:meta}). Ionization signal below the ground-level ionization threshold has been observed in all of the single-pass EII measurements for Mg-like ions~\cite{Montague1983,Belic1987a,Howald1986a,Djuric1993,Hayton1994,Zhang2002,Becker2012}.

Above the threshold for DI of a $3s$ electron the experimental cross section rises monotonically up to about 710~eV, at which point a steeper rise sets in. Up to this energy the theoretical cross section curves from \cite{Dere2007} and \cite{Mitnik1998a} show the same behavior. According to the calculations only DI of a $3s$ valence electron [\eref{eq:DI3s}] contributes to the EII cross section in this energy range. The steep rise at about 710~eV is caused by EA involving $2p\to3p$ and $2p\to3d$ excitations. The level splittings within the excited $2p^5\,3s^2\,3l'$ configurations lead to several closely spaced EA steps in the calculation of Mitnik~\etal~\cite{Mitnik1998a}. There is only one step associated with $2p\to3l'$ EA in the calculation of Dere~\cite{Dere2007} who used a configuration-average approach. This very coarse EA structure is significantly different from the experimental cross section in the 740--810~eV energy range.  At higher energies further EA steps are predicted by the theory which are associated with higher $2p\to n'l'$ excitations [\eref{eq:EA2p}]. In principle there are infinitely many excitation steps converging towards the $2p$ DI threshold as $n'$ approaches infinity. For practical reasons, Mitnik \etal \cite{Mitnik1998a} and Dere \cite{Dere2007} considered $2p\to n'l'$ EA only up to $n'=6$ and $n'=4$, respectively.  The comparison with the present experimental cross section (\fref{fig:EII}) suggests that $2p\to n'l'$ EA cross sections with $n'\geq 4$ are almost insignificant, in contrast to the theoretical predictions. This situation is similar to what has been seen by us in other iron M-shell ions \cite{Hahn2011,Hahn2011a,Hahn2012,Hahn2013a} and has also been investigated and discussed in more detail by Kwon and Savin \cite{Kwon2012} for EII of Fe$^{11+}$.

In contrast to the theoretical prediction, the experimental cross section does not display clear steps. The experimental cross section exhibits REDA [\eref{eq:REDA}] and READI [\eref{eq:READI}] resonances which appear on the low energy side of each corresponding EA threshold and thus blur the EA step structure. Moreover, the amplitudes of EA and REDA processes may destructively interfere, e.g., when $n''l''$ in \eref{eq:REDA} equals $n'l'$ in \eref{eq:EA2p}, leading to additional distortions of EA cross sections. Such interferences have been observed and discussed, e.g., for Li-like C$^{3+}$ \cite{Teng2000a} and O$^{5+}$ \cite{Mueller2000} ions and may also be present here. They can only be accounted for by a unified theoretical treatment as applied in \cite{Teng2000a} and \cite{Mueller2000}. REDA and READI resonances were not included in the more conventional theoretical calculations of Dere~\cite{Dere2007} and Mitnik~\etal~\cite{Mitnik1998a}.

Depending on the electron-ion collision energy, REDA or READI resonances can contribute up to about 30\% of the present total EII cross section of Fe$^{14+}$.  The resonances are less pronounced than in the case of EII of Na-like Fe$^{15+}$ \cite{Linkemann1995a}. For the EII measurements discussed in \sref{sec:introEII}, no REDA or READI resonances were observed for most other ions from the Mg-like isoelectronic sequence~\cite{Montague1983,Belic1987a,Howald1986a,Djuric1993,Hayton1994,Zhang2002} due to limited energy resolution, limited statistical accuracy, and the use of coarse experimental energy grids in these measurements. An exception is the recent study of Becker \etal \cite{Becker2012} who measured the cross section for EII of Ar$^{6+}$ on a fine energy grid with sufficiently low statistical uncertainties for the observation of REDA and READI resonances. These appear to be weaker for Ar$^{6+}$ than for the present more highly charged Fe$^{14+}$.
\subsubsection{Plasma rate coefficient for EII\label{sec:plasmaEII} of Fe$^{14+}$}

\begin{figure}
\includegraphics[width=\figurewidth]{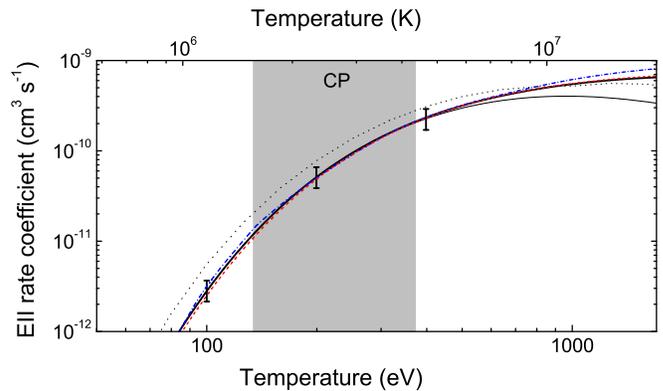}
\caption{\label{fig:plasmaEII} (color online) Rate coefficients for EII of Fe$^{14+}$ in a plasma. The thin and thick solid lines are the present experimentally derived rate coefficients excluding and including the extrapolation by theory, respectively. The error bars correspond to the $\pm26\%$ systematic experimental uncertainty. The short-dashed (red) line is the recommended rate coefficient of Dere \cite{Dere2007}, the short-dash-dotted (blue) line is the rate coefficient of Mitnik~\etal~\cite{Mitnik1998a}, and the dotted curve is the recommended rate coefficient of Arnaud and Raymond \cite{Arnaud1992}. The temperature range where Fe$^{14+}$ is expected to form in a collisionally ionized plasma (CP) is indicated as gray shaded area. It is the same temperature range as in \fref{fig:plasmaDR}.}
\end{figure}

Similar to the procedures applied to DR (\sref{sec:plasmaDR}), we have derived a plasma rate coefficient also for EII of Fe$^{14+}$($3s^2\;^1\!S_0$).  To this end, the experimental cross section has been convoluted by an isotropic Maxwellian electron energy distribution. The resulting plasma rate coefficient  is displayed in \fref{fig:plasmaEII} (thin full line) as a function of plasma temperature.

Also plotted in \fref{fig:plasmaEII} are the theoretical results of Mitnik \etal \cite{Mitnik1998a} and of Dere \cite{Dere2007} who both considered DI and EA but no REDA or READI processes in their calculations. In the temperature range where Fe$^{14+}$ is expected to form in a collisionally ionized plasma both theoretical curves agree with the present experimentally derived result within the experimental systematic uncertainty of $\pm26\%$ at a 90\% confidence limit (\sref{sec:exp}). In the CP temperature range, the recommended rate coefficient of Arnaud and Raymond \cite{Arnaud1992} is 35\%--70\% higher than our present result. This deviation is significantly larger than the 26\% systematic experimental uncertainty. At higher temperatures beyond about $10^7$~K or about 1000~eV, the experimental rate coefficient is significantly lower than the theoretical results. This is largely due to the fact that the present cross section measurement does not extend to electron-ion collision energies beyond 2600~eV, whereas the underlying theoretical cross sections were calculated also at much higher energies.

\begin{table}[b]
\caption{\label{tab:EIIfit}Parameters for the parametrization [\erefs{eq:EIIfit2}{eq:EIIfit4}{--}] of the experimentally derived rate coefficient for EII of Fe$^{14+}$($3s^2\;^1\!S_0$) in a plasma. The parameters are valid in the $6\times10^5 - 9\times10^7$~K temperature range. The numbers in the square brackets denote powers of 10. The value for $E_0$ that is to be used in \erefs{eq:EIIfit3}{eq:EIIfit4}{ and } is 456.2~eV.}
\begin{ruledtabular}
\begin{tabular}{cdcd}
   $i$ & \multicolumn{1}{c}{$a_i$ (cm$^3$\,s$^{-1}$\,eV$^{3/2}$)} &  $i$ & \multicolumn{1}{c}{$a_i$ (cm$^3$\,s$^{-1}$\,eV$^{3/2}$)} \\
  \hline
  0 &   5.17963[-6]  & 5 &   5.68022[-2] \\
  1 &  -1.58627[-5]  & 6 &  -1.09856[-1]  \\
  2 &   1.82163[-4]  & 7 &   1.25272[-1]  \\
  3 &   1.67596[-3]  & 8 &  -7.98108[-2]  \\
  4 &  -1.61240[-2]  & 9 &   2.19428[-2] \\
\end{tabular}
\end{ruledtabular}
\end{table}

In view of the excellent agreement of the theoretical cross section of Dere \cite{Dere2007} with our measured cross section for energies above 1200~eV (\fref{fig:EII}) we used the theoretical cross section of Dere in order to extrapolate our experimental result to energies beyond the upper limit of the experimental energy range at 2600~eV. The convolution of this extrapolated cross section with an isotropic Maxwellian results in our recommended experimentally derived rate coefficient $\alpha_\mathrm{EII}(T)$ for EII of Fe$^{14+}$ in a plasma (thick full line in \fref{fig:plasmaEII}). For the parametrization of $\alpha_\mathrm{EII}(T)$ we use the Burgess-Tully-type scaling from Dere \cite{Dere2007}. Accordingly, the scaled rate coefficient is defined as
\begin{equation}\label{eq:EIIfit1}
\rho = t^{1/2} E_0^{3/2} \alpha_\mathrm{EII}(T)/E_1(1/t)
\end{equation}
with $t = k_BT/E_0$ being the temperature in units of the ionization threshold $E_0=456.2$~eV \cite{Kramida2013}, and with $E_1(1/t)$ denoting the first exponential integral. We have fitted the scaled rate coefficient with a polynomial
\begin{equation}\label{eq:EIIfit2}
\rho\equiv\sum_{i=0}^{9} a_i x^i\mathrm{~~~~with~~~~} x =1 - \frac{\ln 2}{\ln(t+2)}.
\end{equation}
The polynomial coefficients $a_i$ that resulted from the fit are given in \tref{tab:EIIfit}. In the temperature range $6\times10^5 - 9\times10^7$~K, the deviation of the fit from the experimentally derived EII rate coefficient is less than 1\%. The scaled temperature $x$ and the scaled rate coefficient $\rho$ can be inverted to reproduce $T$ and $\alpha_\mathrm{EII}$ using
\begin{equation}\label{eq:EIIfit3}
T = \frac{E_0}{k_B} \left [ \exp \left ( \frac{\ln 2}{1-x} \right ) \right ]
\end{equation}
and
\begin{equation}\label{eq:EIIfit4}
\alpha_\mathrm{EII} = t^{-1/2}E_0^{-3/2}E_1(1/t) \rho.
\end{equation}

\subsection{Comparison between PR and EII\label{sec:comp}}

\begin{figure}
\includegraphics[width=\figurewidth]{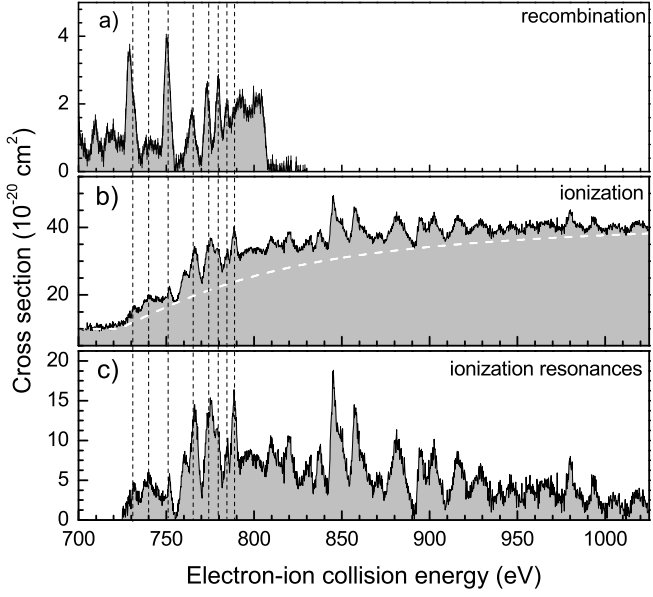}
\caption{\label{fig:comp} Cross sections for recombination and ionization of Fe$^{14+}$ forming Fe$^{13+}$ (a) and Fe$^{15+}$ (b) in the collision energy range of ionization resonances. Panel (c) provides a magnified view of the ionization resonances which has been created by subtraction of the arbitrarily drawn smooth white dashed line in panel (b) from the measured ionization cross section. The vertical dashed lines mark corresponding resonances in the DR and EII spectra.}
\end{figure}

The resonant electron-ion collision processes DR [\erefs{eq:DR0}{eq:DR2s}{--}] and  REDA/READI [\erefs{eq:REDA}{eq:READI}{ and }] involve the population of intermediate multiply excited levels by dielectronic capture (DC) of the initially free electron. For  $2p\to n'l'$ core excitation of Fe$^{14+}$ this process can be written as
\begin{equation}\label{eq:DC2p}
\mathrm{Fe}^{14+}(2p^6\,3s^2) + e^- \rightarrow [\mathrm{Fe}^{13+}]^{**}(2p^5\,3s^2\,n'l'\,nl).
\end{equation}
There are several possibilities for the further decay of the doubly excited intermediate [Fe$^{13+}$]$^{**}$ levels via radiative and autoionizing transitions. DR requires radiative transitions (RT) and net ionization demands the emission of two electrons via autoionization, either sequentially or by a trielectronic interaction where one electron changes into a deeper bound shell and simultaneously transfers sufficient energy to two other bound electrons such that both become ionized. These double autoionization (DA) and auto double (AD) ionization  processes lead to the completion of REDA and READI, respectively. If the intermediate [Fe$^{13+}$]$^{**}$ level autoionizes (AI) by emission of only one electron, then  resonance scattering (RS) will have occurred. The various deexcitation channels can be summarized as
\begin{numcases}{\label{eq:deexci}\!\![\mathrm{Fe}^{13+}]^{**}\rightarrow}
\mathrm{Fe}^{13+} & DR via RT\\
\mathrm{Fe}^{14+}+e^- & RS via AI\\
\mathrm{Fe}^{15+}+2e^- & REDA via DA\\
\mathrm{Fe}^{15+}+2e^- & READI via AD~~~
\end{numcases}

The competition between the various deexcitation channels depends on the relative size of the corresponding transition rates. If the rates $A_\mathrm{RT}$ and $A_\mathrm{DA/AD}$ for RT and DA/AD are of the same order of magnitude, then the corresponding DR and REDA/READI resonances from one and the same intermediate level may be observed in both the measured merged-beams recombination and ionization rate coefficients. These resonances will appear at the same electron-ion collision energy since the resonance energy is determined by the initial resonant DC process. If, however, both rates differ by orders of magnitude, the corresponding resonance will appear only either in the recombination spectrum (if $A_\mathrm{RT}\gg A_\mathrm{DA/AD}$) or in the ionization spectrum (if $A_\mathrm{RT}\ll A_\mathrm{DA/AD}$).

The simultaneous measurement of PR and EII under the same experimental conditions provides a unique opportunity for a comparison of recombination and ionization resonances \cite{Linkemann1995c}. Figure~\ref{fig:comp} shows the measured merged-beams PR and EII cross sections in the energy range where ionization resonances are observed. In \fref{fig:comp}(c) a magnified view of the ionization resonances is presented. It has been obtained by subtraction of an arbitrarily chosen smooth \lq background\rq~cross section [white dashed line in \fref{fig:comp}(b), meant to approximately represent the sum of DI and EA] from the measured EII cross section.

The comparison between PR and EII shows that corresponding recombination and ionization resonances are an exception. Only the $(2p^5\,3s^2\,3d\;^3\!D_1)\;nl$ and $(2p^5\,3s^2\,3d\;^1\!P_1)\;nl$  intermediate levels (see \fref{fig:rec}) contribute to both to DR and REDA/READI. REDA and READI resonances beyond the series limits at 794.95 and 807.29~eV (\tref{tab:Eexci}) are associated with higher $2p\to n'l'$ $(n'\geq4$) core excitations. The corresponding highly-excited intermediate levels strongly autoionize and therefore do not contribute significantly to ($\Delta N > 1$) DR.

\section{Summary and conclusions}\label{sec:summary}

Absolute rate coefficients for DR of Fe$^{14+}$([Ne]\,$3s^2\;^1\!S_0$) ions forming Fe$^{13+}$ and for EII of Fe$^{14+}$([Ne]\,$3s^2\;^1\!S_0$) ions forming Fe$^{15+}$ have been derived from storage ring measurements. The present results together with our previous results for other Fe-M-shell and Fe-L-shell ions \cite{Schippers2010,Hahn2014} (and references therein) are relevant for the modelling of the charge balances in photoionized plasmas (PP) and collisionally ionized plasmas (CP). Moreover, the present results benchmark the most recent theoretical calculations for DR \cite{Altun2007a} and EII \cite{Dere2007} of Fe$^{14+}$. In the relevant temperature ranges (PP and CP ranges for DR, only CP range for EII) both theoretical results agree with the experimentally derived rate coefficients within the systematic experimental uncertainty. On the other hand, the recommended DR and EII rate coefficients of Arnaud and Raymond \cite{Arnaud1992} deviate significantly from the present experimentally derived plasma rate coefficients.

The theoretical work of Mitnik \etal \cite{Mitnik1998a} allowed for detailed comparison between theoretical and experimental EII cross sections. There are considerable differences at certain electron-ion collision energies mainly related to the neglect of REDA/READI processes in the calculation. Nevertheless, the plasma rate coefficient of Mitnik \etal is still in excellent agreement with experiment. The situation is similar for the theoretical cross section of Dere \cite{Dere2007}. Apparently, the theoretical deficiencies are smeared out and mutually cancel when the cross section is convoluted with the plasma electron energy distribution. Certainly, it cannot be expected that this will always be the case. For a comprehensive theoretical understanding of EII of atomic ions, REDA and READI resonances as well as interference effects have to be included in a unified approach that goes beyond the widely used independent-processes approximation.

It should be noted, that the common normalization of our PR and EII results from one and the same experiment constitutes an additional constraint for benchmarking. In addition, the present simultaneous measurement of PR and EII provides a unique opportunity to study the correspondence between recombination and ionization resonances. Corresponding resonances have been observed for a few doubly excited levels associated with $2p\to3d$ core excitation. This implies that radiative and autoionization rates for the decay of these doubly excited $2p^5\,3s^2\,3d\,nl$ intermediate levels are of the same order of magnitude.

\begin{acknowledgments}
We thank the MPIK accelerator and TSR groups for their excellent support during the beam time. Financial support by the Deutsche Forschungsgemeinschaft  (DFG, contract no.\ Schi378/8-1) and by the Max-Planck-Gesellschaft (MPG) is gratefully acknowledged. MH, ON, and DWS were supported, in part, by grants from the NASA Astronomy and Physics Research and Analysis (APRA) Program and the NASA Solar and Heliospheric Physics Supporting Research and Technology Program.
\end{acknowledgments}


%
\end{document}